\documentclass[sigconf]{acmart}
\usepackage{subcaption}
\usepackage{enumitem}
\usepackage{tabularx}
\usepackage{wrapfig}
\usepackage{lipsum}
\usepackage{amsmath}
\usepackage{multirow}
\usepackage{graphicx}
\usepackage{booktabs}
\usepackage{makecell}
\usepackage{algorithm}
\usepackage{algorithmic}
\usepackage{balance} 
\usepackage[normalem]{ulem}

\renewcommand\footnotetextcopyrightpermission[1]{} % removes footnote with conference information in first column
\AtBeginDocument{%
  }

\setcopyright{acmlicensed}
\copyrightyear{2026}
\acmYear{2026}
\acmDOI{XXXXXXX.XXXXXXX}
\acmConference[Conference acronym 'XX]{Make sure to enter the correct
  conference title from your rights confirmation email}{February 12,
  2026}{Hong Kong, CN}
\acmISBN{978-1-4503-XXXX-X/2026/02}

\begin{document}

% \title{Beyond Trial-and-Error: Learning to Compose Reusable \\ Capabilities for Cross-domain Agentic Workflow Design}
\title{Learning to Compose for Cross-domain \\ Agentic Workflow Generation}

\author{Jialiang Wang$^1$, Shengxiang Xu$^3$, Hanmo Liu$^1$$^2$, Jiachuan Wang$^4$, Yuyu Luo$^2$, Shimin Di$^3$, Min-Ling Zhang$^3$, Lei Chen$^1$$^2$}
\email{jwangic@connect.ust.hk, xushx@seu.edu.cn, hliubm@connect.ust.hk, wangjc@slis.tsukuba.ac.jp, yuyuluo@hkust-gz.edu.cn, shimin.di@seu.edu.cn, zhangml@seu.edu.cn, leichen@cse.ust.hk}
\orcid{0009-0005-0850-0389, 0009-0003-4964-8406, 0000-0002-6471-0226, 0000-0001-6473-8221, 0000-0001-9530-3327, 0000-0002-7394-0082, 0000-0003-1880-5918, 0000-0002-8257-5806}
\affiliation{%
  \institution{%
  \textsuperscript{1}Hong Kong University of Science and Technology, Hong Kong SAR, China\\
  \textsuperscript{2}Hong Kong University of Science and Technology (Guangzhou), Guangzhou, China\\
  \textsuperscript{3}Southeast University, Nanjing, China\\
  \textsuperscript{4}University of Tsukuba, Tsukuba, Japan}
  \city{}
  \country{}
}
\renewcommand{\shortauthors}{Jialiang Wang et al.}

\begin{abstract}
Automatically generating agentic workflows---executable operator graphs or codes that orchestrate reasoning, verification, and repair---has become a practical way to solve complex tasks beyond what single-pass LLM generation can reliably handle.
Yet what constitutes a good workflow depends heavily on the task distribution and the available operators.
Under domain shift, current systems typically rely on iterative workflow refinement to discover a feasible workflow from a large workflow space, incurring high iteration costs and yielding unstable, domain-specific behavior.
In response, we internalize a \emph{decompose-recompose-decide} mechanism into an open-source LLM for cross-domain workflow generation.
\emph{To decompose}, we learn a compact set of reusable workflow capabilities across diverse domains.
\emph{To recompose}, we map each input task to a sparse composition over these bases to generate a task-specific workflow in a single pass.
\emph{To decide}, we attribute the success or failure of workflow generation to counterfactual contributions from learned capabilities, thereby capturing which capabilities actually drive success by their marginal effects.
Across stringent multi-domain, cross-domain, and unseen-domain evaluations, our 1-pass generator surpasses SOTA refinement baselines that consume 20 iterations, while substantially reducing generation latency and cost.

\end{abstract}

% %%
% %% The code below is generated by the tool at http://dl.acm.org/ccs.cfm.
% %% Please copy and paste the code instead of the example below.


% %% A "teaser" image appears between the author and affiliation
% %% information and the body of the document, and typically spans the

\received{12 February 2026}
\received[revised]{12 February 2026}
\received[accepted]{12 February 2026}

\maketitle

\section{Introduction}
\label{sec:introduciton}

\begin{figure*}[t!]
    \vspace{-5px}
    \centering
    \includegraphics[width=0.86\textwidth]{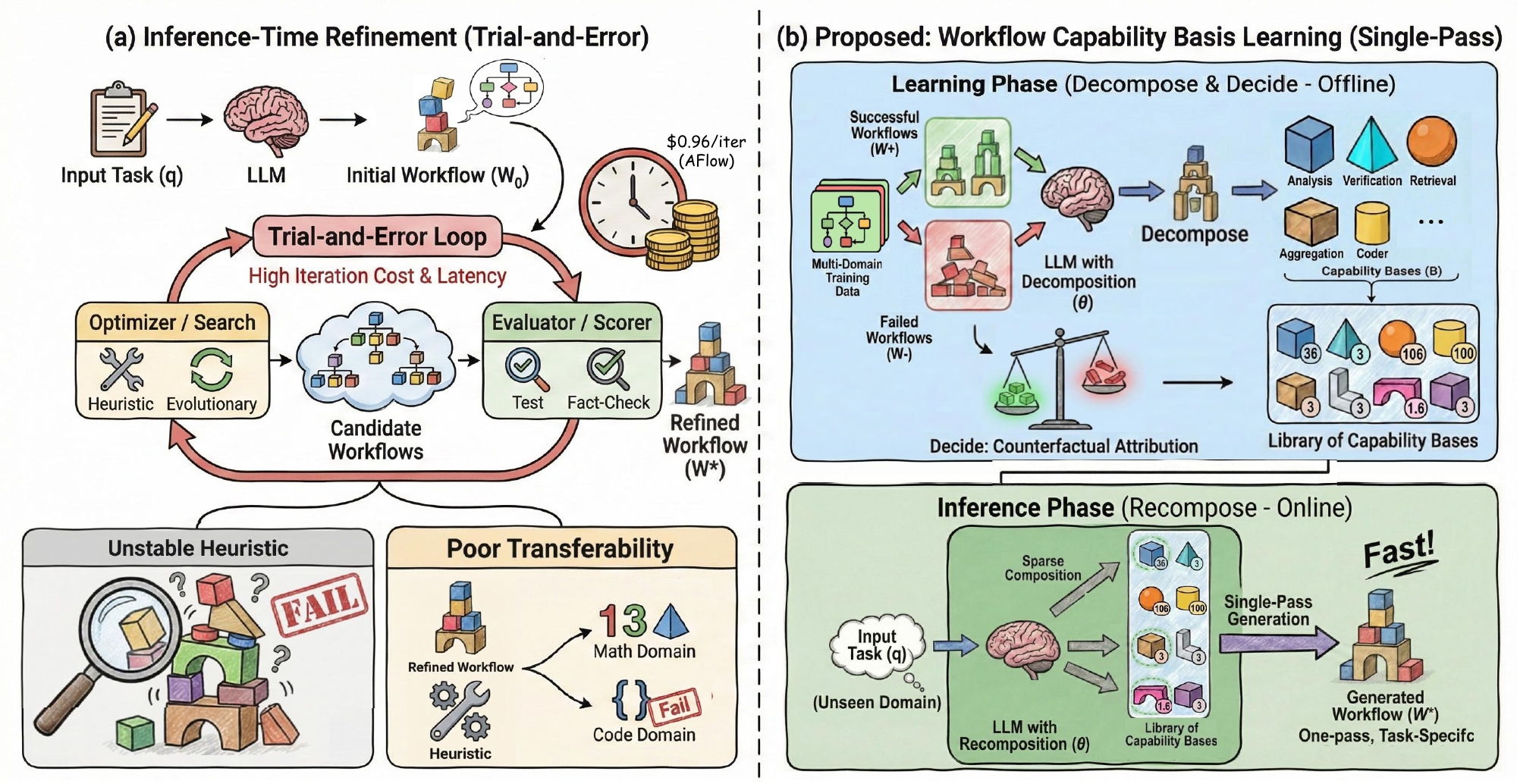}
    \vspace{-12px}
    \caption{Two workflow generation paradigms. Left: inference-time refinement resorts to trial-and-error in a large workflow space. Right: CapFlow internalizes ``\emph{decompose-recompose-decide}'' into LLMs, enabling single-pass generation across domains.}
    \label{fig:introduction_figure}
    \vspace{-6px}
\end{figure*}

Large language models (LLMs) have demonstrated strong zero-shot capabilities in open-domain question answering and code generation~\cite{brown2020language,chen2021evaluating,ouyang2022training}.
Yet, these capabilities are initially realized through single-pass generation: the model produces a final answer or program in one shot.
For complex tasks, single-pass generation often hits a structural ceiling~\cite{hong2023metagpt,chen2024agentverse}.
Beyond being correct, solutions must satisfy constraints, admit external tools, support error correction, and remain reliable under tight latency and cost budgets.

%介绍agentic workflow的概念和operator案例
To push beyond this ceiling, agentic workflows~\cite{yao2022react,zhuge2024gptswarm,liu2024dynamic} have emerged as a practical approach after the Chain-of-Thought~\cite{wei2022chain}. 
By making the solving procedure more explicit, an agentic workflow decomposes the task into an ordered, executable composition of operators and executes them under a control structure.
Concretely, workflow generation for complex tasks requires deciding which operators to invoke and how to compose them into a topology (e.g., a sequence, a branching graph) that determines how intermediate states are produced and consumed~\cite{yemas}.
For example, open-domain reasoning benefits from operators that retrieve evidence and compare multiple aspects~\cite{wangself}; mathematical problems require verification and counterexample-checking operators~\cite{cobbe2021training}; and code tasks demand the topology of a test-and-repair loop~\cite{madaan2023self}.

Recent years have seen rapid advances in building agentic systems with multi-agent workflows.
Pioneering works~\cite{hong2023metagpt,chen2024agentverse} use manually designed operator pipelines or collaboration structures---effective in specific settings but typically fixed regardless of the input task, thereby limiting adaptivity and generality.
Motivated by the cost and brittleness of manual design, recent work has moved toward automating workflow generation, aiming to reduce human engineering and tailor workflows to the task.
For example, systems such as AFlow~\cite{zhangaflow} and related automated frameworks~\cite{huautomated,wang2025scoreflow,xu2025robustflow} seek to generate or refine workflows with minimal manual specification, improving scalability across diverse domains.

%%指出agentic workflow在inference上的技术机理
A common strategy behind these automated systems is to place workflow generation inside the inference loop (shown in Fig.~\ref{fig:introduction_figure}(a)).
Given a task, the model first generates one or more candidate workflows, and then samples and improves them via search or iterative refinement---through best-of-N sampling~\cite{khattab2024dspy}, self-reflection rewriting~\cite{wang2025scoreflow}, heuristic structure edits~\cite{zhangaflow}, or evolutionary procedures~\cite{novikov2025alphaevolve} that repeatedly select, mutate, and re-evaluate workflows.
Overall, this workflow refinement paradigm treats workflow generation as a trial-and-error inference over a large workflow space, trading high iteration costs for effectiveness and generality.

However, the fact that workflows can raise the ceiling does not imply that an LLM can naturally generate effective, task-specific workflows for diverse domains.
In fact, the past paradigm resembles attaching an external optimizer at inference time rather than endowing the model with transferable workflow generation capability.
Fig.~\ref{fig:introduction_figure}(a) exposes two limitations in handling domain shift.
First, workflow generation is frequently driven by LLM-only heuristics~\cite{huautomated} or stochastic refinement~\cite{zhangaflow}.
This limited stability and controllability offer no guarantee about workflow quality when task distribution shifts.
Second, good workflow criteria and strategies are difficult to standardize across domains.
Workflow criteria and heuristics that work in one domain may fail or even backfire in another~\cite{zhang2025gnns,trirat2025agentic}, yielding pronounced generalization variance.
Therefore, workflow generation often amounts to within-task trial-and-error to approach a feasible workflow at inference time, rather than \emph{learning a transferable mapping from task semantics to workflow structure decisions}.
We further characterize this missing mechanism through two coupled gaps:
(1) a capability decomposition gap: LLMs often represent tasks at the content level but lack a representation that directly exposes which workflow-relevant capabilities are needed; and
(2) a capability recomposition gap: even when useful workflow patterns are known, the model lacks a controllable way to select and combine the right capabilities for a new task.

Notably, these gaps do not imply that workflow generation is wholly domain-specific.
Our study in Sec.~\ref{sec:methodology_bases} shows that, despite task-surface differences across domains, many successful workflows repeatedly instantiate similar underlying capability factors (e.g., multifaceted analysis, verification/repair, and aggregation).
As illustrated in Fig.~\ref{fig:introduction_figure}(b), if a model could represent these capability factors in a parametric, reusable form and compose them on demand for new tasks, cross-domain workflow generation could shift from trial-and-error inference to a single-pass structural decision.

% \footnote{
% [Fixed] \# shimin: 1) where is ``decompose-recompose-decide'',
% 2) be simple and straightforward
% }
% {\color{blue}
To bridge the decomposition and recomposition gaps, we internalize a ``\emph{decompose-recompose-decide}'' mechanism into the open-source LLM so that workflow generation does not rely on heuristic trial-and-error inference.
\emph{To decompose}, we learn a compact set of reusable capability bases across diverse domains, capturing recurring workflow factors that generalize in the latent space.
\emph{To recompose}, we map an input task to a sparse composition over these bases, providing a controllable way to reuse capabilities for new tasks.
\emph{To decide}, we attribute the success or failure of workflow generation to counterfactual contributions from learned capabilities, capturing which capability bases truly contribute to workflow generation through their marginal effects under domain shifts.
The resulting model generates the executable, task-specific agentic workflow in a single pass, thereby avoiding costly refinement during inference.
% }
Our contributions are:
\begin{itemize}[leftmargin=*]
    \item We propose Workflow Capability Basis Learning, reframing cross-domain workflow generation from trial-and-error inference into a learnable problem of capability decomposition and recomposition, enabling single-pass, task-specific workflow generation.
    \item We present a workflow generalization framework, CapFlow, centered on shared capability bases and task-conditioned composition, achieving structured decision-making and compositional transfer without replacing the underlying base model.
    \item We curate multi-domain successful and failed workflow data to construct counterfactual contribution attribution and couple it with preference-driven supervision, aligning basis learning with factors that drive success and improving generalization.
    \item Empirical results across multi-domain, cross-domain, and unseen-domain settings show that our 1-pass generation method can exceed inference-time refinement baselines that consume 20 iterations, substantially reducing workflow generation cost.
\end{itemize}

\section{Related Work}
\label{sec:related_work}

\subsection{Automated Agentic Workflows}
\label{sec:related_work_workflows}
% \footnote{
% [Fixed] \# shimin: to save space, be simple for related work
% }
% \footnote{
% \color{red}
% \# shimin: better reorganize to introduce some important concepts of this work:
% (1) training and inference of agentic workflow generation,
% (2) inference strategies of existing works

% \# jialiang: the narrative flow is: agentic workflow (basic concept, manual workflow) -> autoamted workflow generation [two ways: (1) automated workflow refinement (inference); (2) learning to generate workflows (training)]
% }
% \footnote{
% [Fixed] \# shimin: other concepts can be introduced accordingly:
% 	(1) workflow formulation (code/graph/...),
% 	(2) operator
% 	(3) query, task
% 	...
% }
\noindent
\textbf{Agentic workflows} push beyond single-pass answer generation of LLMs by making the solution procedure explicit and executable~\cite{hong2023metagpt,chen2024agentverse}.
A typical workflow maps a user query/task $q$ to an executable composition of \emph{operators} (e.g., multifaceted analysis, verification, aggregation) with a specific \emph{topology} (e.g., sequences, branching graphs, or repair loops), often maintaining intermediate states and tool invocations.
In practice, workflows are instantiated as plans~\cite{qiaobenchmarking}, graphs~\cite{zhang2025gnns}, neural networks~\cite{liu2023dynamic}, or programs~\cite{huautomated}.
However, early approaches often rely on either (i) a manually curated library of workflow templates~\cite{hong2023metagpt,chen2024agentverse} or (ii) a generic planning strategy reused across tasks~\cite{wei2022chain,wangself,madaan2023self}.
These methods quickly face scalability limits as the space of operator choices, topologies, and prompt strategies grows combinatorially, motivating two lines of research on automated workflow generation.
% A growing body of work has explored agentic workflows as a model-agnostic way to push beyond the task-solving ceiling of single-pass LLM generation by making the solution procedure explicit and executable~\cite{hong2023metagpt,chen2024agentverse}.
% Typical workflows decompose problem-solving into purposeful operators---e.g., analysis, candidate generation, verification/testing, aggregation, and repair---and often couple LLM calls with intermediate states and tool invocations.
% {\color{red}
% Many systems represent workflows as plans~\cite{qiaobenchmarking}, graphs~\cite{zhang2025gnns}, neural networks~\cite{liu2023dynamic}, or lightweight programs~\cite{huautomated} executed by an orchestration layer that routes information across steps and triggers tools when needed.
% Despite this success, early approaches commonly assume either (i) a manually curated library of workflow templates~\cite{hong2023metagpt,chen2024agentverse} or (ii) a generic planning strategy~\cite{wei2022chain,wangself,madaan2023self} reused across tasks, with task specificity delegated to prompt engineering and ad hoc orchestration.
% }
% Such assumptions quickly run into scalability limits: the space of plausible workflows grows combinatorially with operator choices, control-flow structures, and prompt strategies, which motivates recent research on automated workflow generation.

\begin{figure*}[!t]
\vspace{-5px}
\centering
\begin{subfigure}[t]{0.60\textwidth}
    \centering
    \includegraphics[width=\linewidth]{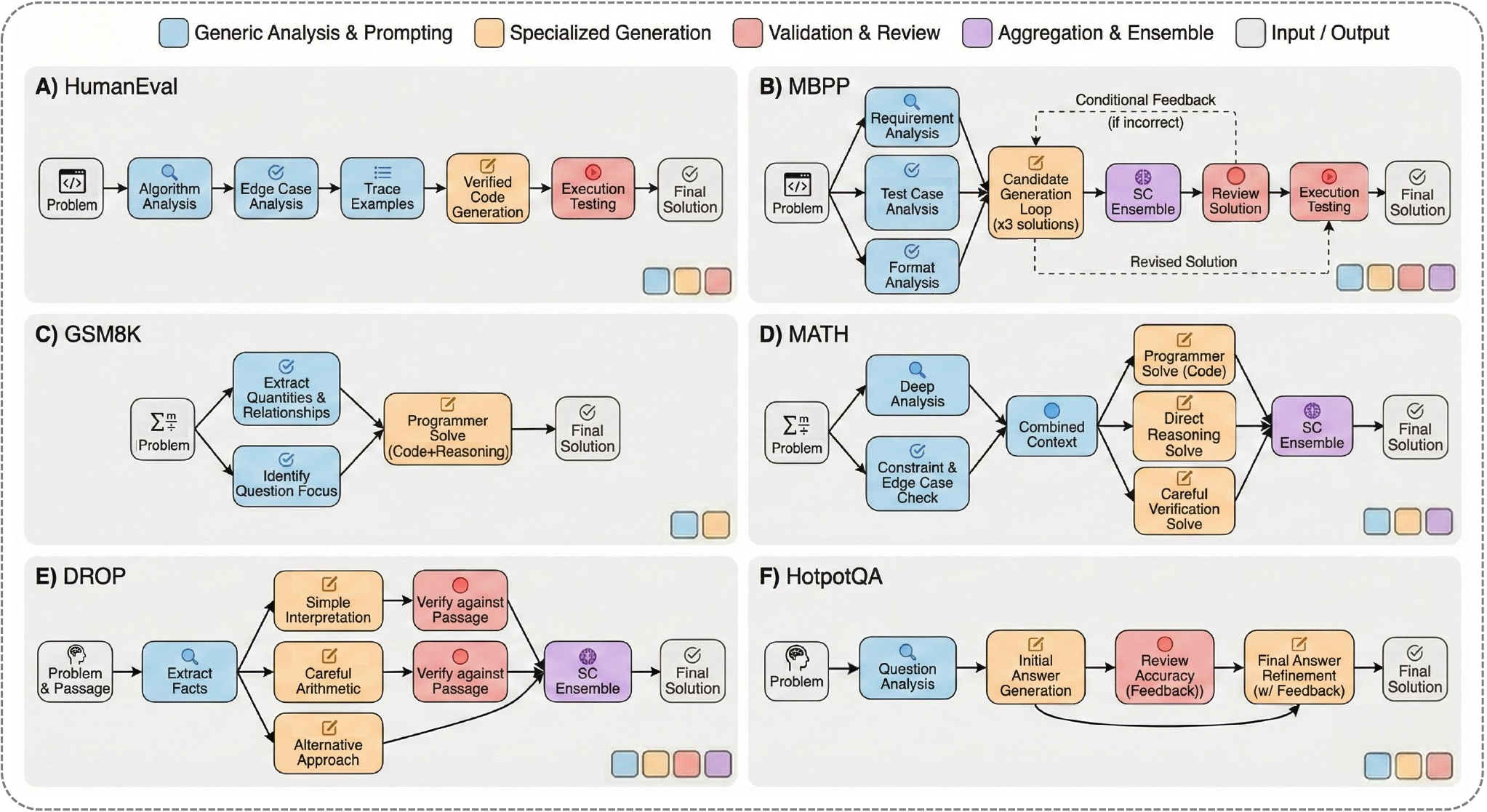}
    \caption{Our 6 most successful agentic workflows for coding, math, and reasoning domains.}
    \label{fig:golden_workflow}
\end{subfigure}
\hfill
\begin{subfigure}[t]{0.38\textwidth}
    \centering
    \includegraphics[width=\linewidth]{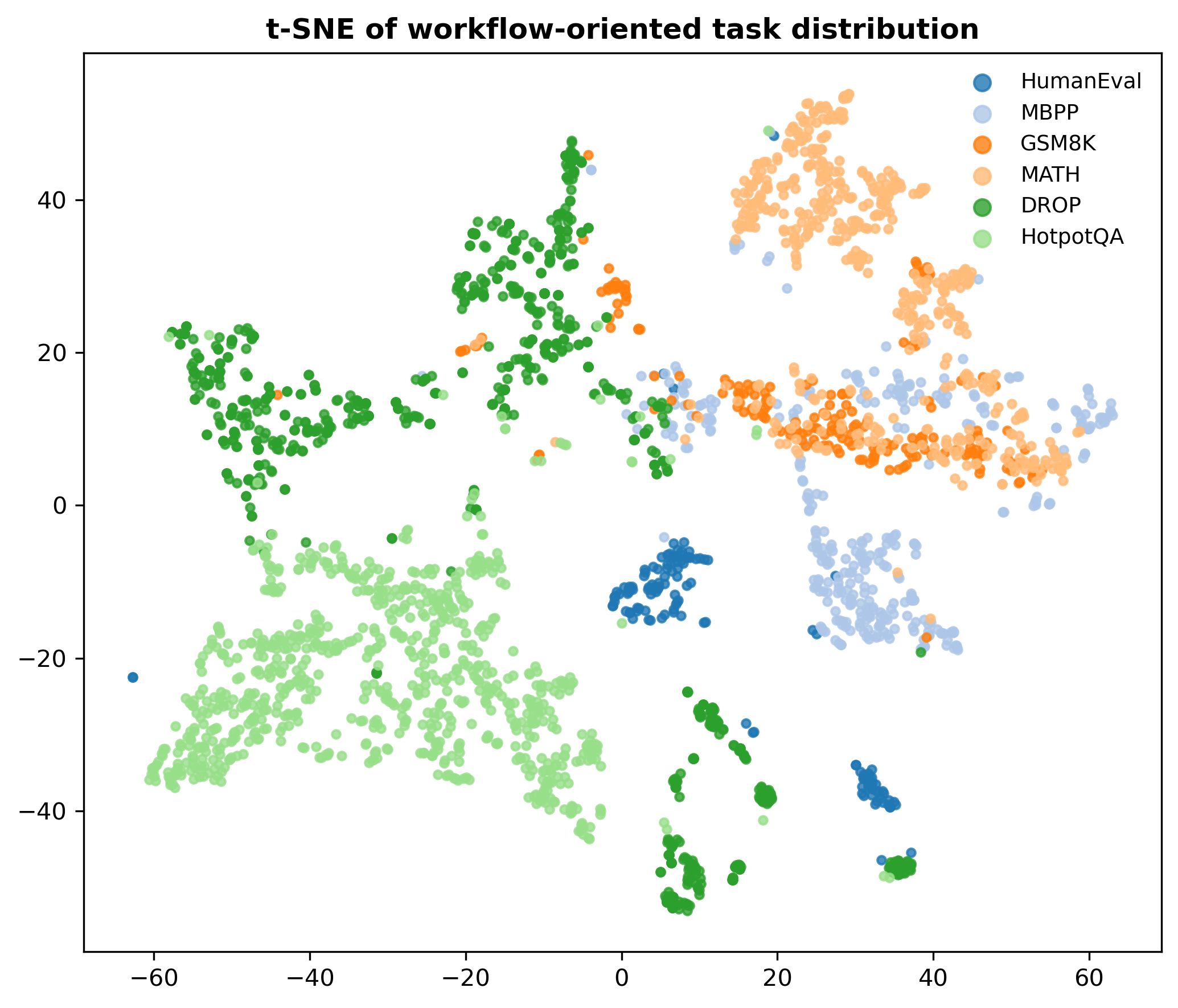}
    \caption{Workflow-oriented task distribution (t-SNE).}
    \label{fig:tsne_by_domain}
\end{subfigure}
\vspace{-10px}
\caption{Cross-domain agentic workflow analysis: (left: structural analysis) highest-success workflows per domain; (right: latent analysis) t-SNE visualization of tasks embedded by learned workflow capabilities.}
\label{fig:workflow_analysis}
\vspace{-5px}
\end{figure*}

\noindent
\textbf{Automated Workflow Refinement.}
A prevalent paradigm is to construct an external optimizer over workflow space $\mathcal{W}$ and treat workflow generation as an inference-time optimization problem:
\begin{equation}
W^{*}\;=\;\arg\max_{W\in\mathcal{W}}\;G(W;\,q)
\end{equation}
In this paradigm, the model samples candidate workflows $W$ and then improves them via search or iterative refinement to maximize the evaluation function $G$ for the given task $q$.
For example, AFlow~\cite{zhangaflow} performs search over code-represented workflows with LLM-in-the-loop refinement and Monte Carlo tree search, while GPTSwarm~\cite{zhuge2024gptswarm} optimizes graph-structured multi-agent topologies by editing nodes/edges.
ADAS~\cite{huautomated} and MASS~\cite{zhou2025multi} also advocate treating agentic system construction as search over modular building blocks and compositions.
Despite strong performance, this trial-and-error inference can be costly and domain-specific:
(i) it incurs substantial and hard-to-budget workflow execution cost at each iteration with diminishing returns~\cite{zhangaflow};
(ii) workflow generation is often driven by LLM-only heuristics or stochastic edits, yielding limited controllability and unstable performance outcomes~\cite{xu2025robustflow};
and (iii) effective workflow criteria and editing heuristics are difficult to transfer across different domains, leading to pronounced generalization variance under domain shift~\cite{zhang2025gnns,trirat2025agentic}.

\noindent
\textbf{Learning to Generate Workflows.}
Beyond heuristic refinement, several works learn decision policies for agentic systems from data $\mathcal{D}$ by optimizing a workflow generator $p_\theta(W\mid q)$, viewing workflows as trajectories or code whose quality is judged by task success:
\begin{equation}
\theta^{*} \;=\;\arg\max_{\theta}\;\mathbb{E}_{q\sim\mathcal{D}}\;\mathbb{E}_{W\sim p_{\theta}(\cdot\mid q)}\!\left[\,G(W;\,q)\,\right]
\end{equation}
This learning-based paradigm covers supervised fine-tuning or behavioral cloning to imitate high-quality workflows~\cite{hu2022lora}, preference-based learning that exploits relative judgments among candidate workflows~\cite{rafailov2023direct}, and reinforcement-style optimization when feedback is available.
For example, ScoreFlow~\cite{wang2025scoreflow} leverages direct preference optimization to exploit quantitative feedback.
FlowReasoner~\cite{gao2025flowreasoner} trains a query-level meta-agent with reinforcement learning to generate personalized multi-agent systems.
MAS-GPT~\cite{yemas} builds query-workflow pairs via a consistency-oriented data construction pipeline to train workflow generation.
This learning-based perspective offers a principled alternative to handcrafted heuristics by grounding workflow generation in data.
However, two limitations remain salient for cross-domain transfer:
(i) naive imitation can be brittle under task distribution shift~\cite{de2019causal}, since it \emph{does not isolate which workflow factors causally drive success};
(ii) learned behaviors are often entangled in monolithic parameters, \emph{lacking a controllable mechanism for compositional transfer}~\cite{pfeiffer2021adapterfusion} when new tasks require recombining familiar factors in novel ways.
Our work aligns with this learning-based paradigm, while explicitly targeting controllable recomposition for transferable workflow generation under domain shift.

% In this paradigm, behavioral cloning and supervised fine-tuning~\cite{hu2022lora} can capture recurring planning patterns, while preference-based learning~\cite{rafailov2023direct} incorporates relative judgments between candidate solutions.
% For example, ScoreFlow~\cite{wang2025scoreflow} proposes a gradient-based optimization framework in continuous space and introduces direct preference optimization to exploit quantitative feedback.
% FlowReasoner~\cite{gao2025flowreasoner} trains a query-level meta-agent with reinforcement learning to generate personalized multi-agent systems conditioned on a query.
% MAS-GPT~\cite{yemas} trains LLMs to build multi-agent systems using a consistency-oriented data construction pipeline over query-MAS pairs.
% This learning-based perspective offers a principled alternative to hand-designed heuristics by grounding workflow design in data.
% Nevertheless, directly applying trajectory learning to workflow design faces two challenges that our formulation makes explicit. 
% First, purely imitating successful trajectories can be brittle under distribution shift~\cite{de2019causal}, since it does not isolate which aspects of a workflow are causally responsible for success, making learned behaviors fragile in unseen domains.
% Second, even when some reusable patterns are learned, models often lack a controllable mechanism for compositional transfer~\cite{pfeiffer2021adapterfusion}: unseen tasks may require recombining familiar skills in novel ways, and naive fine-tuning tends to entangle these skills in monolithic parameters.

\subsection{Parameter-efficient and Modular Adaptation}
\label{sec:related_work_peft}
Parallel to workflow-centric systems, substantial progress has been made in parameter-efficient LLM adaptation, including low-rank parameterizations~\cite{hu2022lora} that enable specialization without full fine-tuning.
Relatedly, modular and conditional-computation approaches route inputs to a subset of parameters, often framed as mixtures of experts~\cite{cai2025survey} or mixtures of adapters~\cite{wu2023mole,zhong2024multi}, with connections to low-dimensional predictive subspaces~\cite{ando2005framework,argyriou2006multi}.
Crucially, however, most prior work on routing and modular adaptation targets efficient computation, capacity scaling, or multi-task specialization, and many multi-domain methods use explicit domain identifiers to drive discrete specialization~\cite{dai2021generalizable,zhang2024m3oe}.
In contrast, our problem requires learning a structured decision mechanism aligned with executable workflow success under domain shift.
We interpret modularity not as an engineering device for compute allocation, but as a vehicle for capability recomposition: the routed components correspond to reusable workflow capability factors, and the routing decision is trained to dominate single-pass workflow generation.
This distinction is central to our work: we learn a ``\emph{decompose-recompose-decide}'' policy that captures transferable structural decision rules, rather than performing predictive fusion across experts.

\section{Preliminary}
\label{sec:preliminary}

\noindent
\textbf{Motivational Study.}
To validate our motivation for reusable workflow capabilities, we conduct both \emph{structural} and \emph{latent} analyses of successful workflows across domains.
We visualize the most successful workflows from each domain in our dataset in Fig.~\ref{fig:golden_workflow} and analyze their operators and topologies.
We observed recurring workflow patterns such as multifaceted analysis (blue nodes) and ensemble aggregation (purple nodes) across domains (B, D, E).
In contrast, domain-specific behaviors mainly arise from specialized operators, e.g., code testing for coding tasks (red nodes in A, B) and multi-perspective task solving for QA tasks (orange nodes in E).

% To validate our key motivation for reusable workflow capabilities, we conduct surface and latent studies of successful workflow patterns across domains.
% We extract the most successful workflows from our multi-domain dataset and analyze their operators and structures.
% Figure~\ref{fig:golden_workflow} visualizes the agentic workflow designs with the highest success rates from each domain.
% We observed shared workflow patterns such as task decomposition (blue nodes) and ensemble aggregation (purple nodes) across domains (B, D, E).
% Domain-specific designs mainly emerge from specialized operators, such as code testing in coding tasks (red nodes in A, B) and multi-perspective reasoning in QA tasks (orange nodes in E).

% \footnote{
% [Fixed] \# shimin: may need to explain more simple and clearly
% }
To further examine the significance of aligning the task semantics to workflow behaviors, we embed each task using the learned workflow capabilities and visualize the resulting workflow-oriented task distribution via t-SNE in Fig.~\ref{fig:tsne_by_domain}.
Without using any domain identifier in learning, tasks from the same domain naturally cluster together, suggesting that (1) \emph{same-domain tasks often require similar workflow capabilities.}
Meanwhile, tasks from different domains can overlap in the workflow space, indicating that (2) \emph{successful workflows often share latent factors despite surface task differences.}
These observations motivate explicitly modeling reusable workflow capabilities and recomposing them for new tasks.

% To further demonstrate the significance of aligning the task semantics to workflow design, we embed each task using learned workflow capabilities and visualize the workflow-oriented task distribution using t-SNE in Figure~\ref{fig:tsne_by_domain}.
% Without using any domain identifier, tasks from the same domain naturally cluster together, indicating that (1) \emph{same-domain tasks often require similar workflow capabilities.}
% Surprisingly, workflows from different domains may overlap in the embedding space, suggesting that (2) \emph{successful workflows often share latent factors despite surface task differences.}
% This observation motivates us to explicitly model these shared factors as reusable capability bases, enabling cross-domain recomposition.

\noindent
\textbf{Problem Setup and Data.}
Each open-domain task contains a question text $q$ and a formatted prompt (including instruction and operator descriptions).
The target output is an executable workflow program $W$ that contains both (i) workflow topology (control flow) and (ii) customized operator prompts.
Since existing benchmarks rarely provide executable workflow programs together with both success and failure outcomes, we curate a multi-domain dataset containing 180+ unique workflows with performance records over 6 datasets from 3 distinct domains (coding, math, and reasoning):
% \footnote{
% [Fixed] \# shimin: 1) one of core contributions?
% 2) why we need to curate data sets?
% }
\begin{equation}
\mathcal{D}=\{ (q_i,\{ (W_j, s_{i,j})\} _{j=1}^{N_w})\} _{i=1}^{N_q},\quad s_{i,j}\in[0,1]
\end{equation}
where $s_{i,j}$ indicates whether the workflow solves the task.
We formulate workflow generation as conditional code generation:
\begin{equation}
W \sim p_{\Theta(q)}(W\;|\;x(q))
\label{eq:workflow_generation}
\end{equation}
where $x(q)$ is the formatted prompt instruction, and $\Theta(q)$ denotes task-conditioned model parameters induced by our capability composition mechanism.

\section{Methodology}
\label{sec:methodology}
The existing trial-and-error paradigm does not produce transferable decision rules---hence the cost explosion, instability, and cross-domain variance discussed in Sec.~\ref{sec:related_work_workflows}.
We instead ask a different question: \emph{Can workflow generation be turned into a learned decision mechanism that maps task semantics to customized workflow structure in a single pass?}
There are two key challenges behind:
(i) a \emph{capability decomposition gap}---tasks are not naturally represented in a space aligned with effective workflow factors; and
(ii) a \emph{capability recomposition gap}---even if reusable patterns exist, the model lacks a controllable way to recombine them under domain shift.

To bridge these gaps, we internalize a ``\emph{decompose-recompose-decide}'' mechanism into the model by 
(1) introducing a small set of reusable capability bases over latent workflow factors and 
(2) training a task-conditioned capability composer that selects a sparse combination of these bases for each new task.
CapFlow is a single-pass generator that generates executable, task-specific agentic workflow programs without iterative refinement at inference time.

\begin{figure*}[t!]
    \vspace{-5px}
    \centering
    \includegraphics[width=0.9\textwidth]{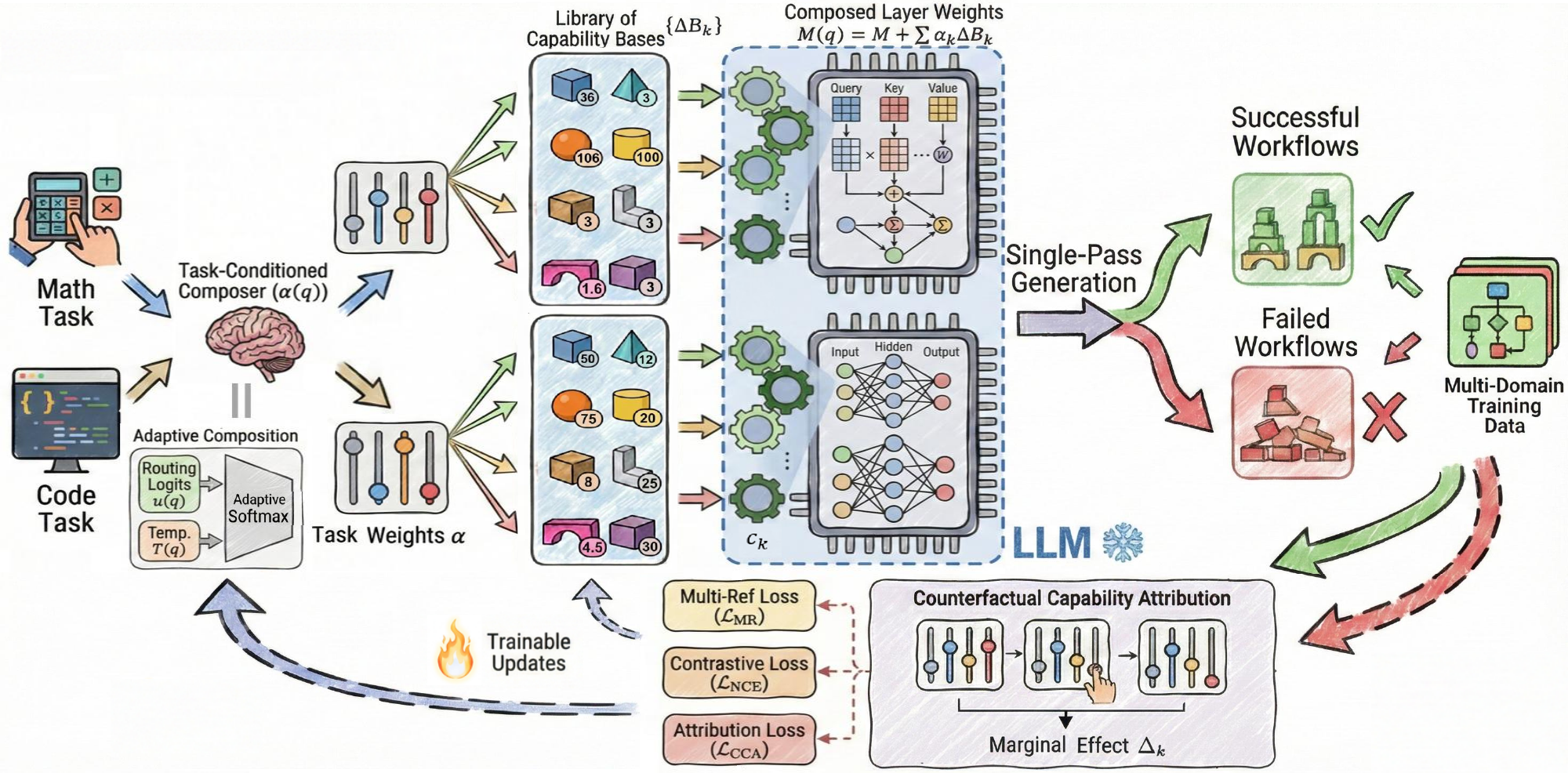}
    \vspace{-10px}
    \caption{Overview of workflow capability composition (CapFlow): the task-conditioned composer decomposes each query into a sparse mixture of reusable capability bases, steering the LLM toward successful workflows and away from failures.}
    \label{fig:methodology_figure}
    \vspace{-5px}
\end{figure*}

\subsection{Workflow Capability Bases}
\label{sec:methodology_bases}
% \textbf{Capability Basis.}
We operationalize the findings in Sec.~\ref{sec:preliminary}---that many good workflows share latent capability factors despite surface-level task differences---by parameterizing these factors as a compact set of reusable \emph{capability bases}.
Concretely, starting from a frozen weight LLM $f_{\theta_0}$, we introduce a small set of $K$ lightweight adaptation bases into a chosen subset of linear transformations (e.g., attention and MLP projections).
Each basis serves as a reusable \emph{latent workflow factor} that can bias the model toward certain operator choices, control-flow patterns, and prompt-writing behaviors that repeatedly emerge in successful workflows.

Formally, consider a linear transformation with weight matrix $M \in \mathbb{R}^{d_{\text{out}}\times d_{\text{in}}}$.
We augment it with $K$ rank-$r$ basis updates:
\begin{equation}
\Delta B_k \;=\; c_k \, U_k V_k^\top,\quad
U_k\in\mathbb{R}^{d_{\text{out}}\times r},\;
V_k\in\mathbb{R}^{d_{\text{in}}\times r},\;
c_k\in\mathbb{R}_{+},
\end{equation}
where $U_k$ and $V_k$ parameterize a capability basis and $c_k$ is a learnable per-basis scale.
These parameters are module-specific but shared across all tasks.
Intuitively, task-specific workflow generation emerges from \emph{selectively composing} these bases.

\subsection{Task-Conditioned Capability Composer}
\label{sec:methodology_composer}
Capability bases provide a workflow-aligned representation space, but we still need a mechanism that \emph{maps task semantics into that space} and \emph{selects how to recombine factors} under domain shift.
We model this as a task-conditioned capability composer:
\begin{equation}
\alpha(q) \;=\; g_{\psi}\big(z(q)\big),
\label{eq:controller}
\end{equation}
where $z(q)$ is a task embedding and $g_\psi$ outputs mixture weights over bases.
This design makes the composer depend on the \emph{pure task} rather than on accidental artifacts of the LLM heuristic, which empirically stabilizes training and mitigates hallucinations.

\noindent
\textbf{Adaptive Composition.}
A core difficulty in composition is trading off specialization and reuse.
If composition collapses too early, a few bases dominate and the mechanism stops being compositional; if composition is too diffuse, the model behaves like a weak ensemble and loses controllability~\cite{lepikhingshard}.
To address this, we design a composer that predicts (a) composition logits and (b) a per-task temperature adjustment with coupled multi-layer perceptrons $h_\psi(\cdot)$ and $\Delta t_\psi(\cdot)$:
\begin{equation}
u(q) = h_\psi(z(q))\in\mathbb{R}^K,\qquad
T(q)=\exp\!\left(\log T_0 + \Delta t_\psi(z(q))\right),
\end{equation}
and produces probabilities via
\begin{equation}
% \alpha(q) \;=\; \mathrm{Softmax}\!\left(\ell(q)/T(q)\right)\in\Delta^K,
\alpha(q) \;=\; \mathrm{Softmax}\!\left(\frac{u(q)}{T(q)}\right)\in\Delta^{K-1},
\label{eq:routing}
\end{equation}
Intuitively, $T(q)$ controls how decisive the composer is: tasks that admit multiple workflow capabilities can retain higher entropy, while tasks with clear patterns can yield sharper compositions.

\subsection{Workflow Capability Decomposition}
\label{sec:methodology_decomposition}
Combining the capability bases (Sec.~\ref{sec:methodology_bases}) and the task-conditioned composer (Sec.~\ref{sec:methodology_composer}), we induce a task-specific parameterization of the agentic workflow generator.
Given a task $q$, the weight of each adapted linear transformation is composed as:
\begin{equation}
M(q)\;=\; M + \sum_{k=1}^K \alpha_k(q)\, \Delta B_k.
\label{eq:composed_weight}
\end{equation}
Here $\alpha(q)\in\Delta^{K-1}$ is a probability simplex vector produced by a task-conditioned composer.
Crucially, the same $\alpha(q)$ is broadcast to \emph{all} adapted layers, so that a global task-level capability decision coherently shapes the entire workflow generator (operators and topologies) rather than producing layer-wise, uncoordinated gating.
To encourage \emph{recompositional} behavior and keep inference efficient, we use a sparse instantiation of Eq.~\eqref{eq:composed_weight} by activating only the top-$m$ bases per task and renormalizing their weights, which provides a controlled inductive bias toward few-factor workflow generations (consistent with the empirical patterns in Fig.~\ref{fig:golden_workflow}).
Overall, this mechanism addresses the capability decomposition gap: instead of forcing the model to costly refine an entire workflow structure from scratch for each task, we represent tasks in a space where the relevant capability factors are explicitly \emph{addressable} and \emph{composable}.

\subsection{Learning Basic Workflow Capabilities}
\label{sec:methodology_learning}
Beyond Sec.~\ref{sec:related_work_workflows}, training purely by imitation is ill-suited for workflow generation because (i) each task often admits multiple correct workflows, and (ii) what matters is not stylistic similarity to a reference but \emph{whether the workflow succeeds}.
Our dataset fills this missing supervision with both successful and failed workflows for each task.
We leverage it to learn (a) capability bases that increase the likelihood of successful workflow code and (b) a composer whose task decomposition aligns with success \emph{causally} rather than correlationally (Sec.~\ref{sec:methodology_recomposition}).

Let $W$ denote the workflow program (code) and $x(q)$ the formatted input.
We use the length-normalized causal LM log-likelihood over target tokens $w_t$ to avoid bias toward shorter references:
\begin{equation}
\ell(W,x(q))=\frac{1}{|W|}\sum_{t} \log p_{\Theta(q)}(w_t \mid x, w_{<t}),
\end{equation}

\noindent
\textbf{Multi-reference Success Likelihood.}
For a task $q_i$, let $\mathcal{P}_i$ denote the set of successful workflows and $\mathcal{N}_i$ the set of failures.
Since multiple workflows may succeed, we do \emph{not} force the model to imitate a single reference.
Instead, we maximize the total probability mass assigned to the successful set via a multi-reference objective:
\begin{equation}
\mathcal{L}_{\mathrm{MR}}
\;=\;
-\sum_{i}
\log \sum_{W\in\mathcal{P}_i} \exp\big(\ell(W, x(q_i))\big).
\label{eq:mr}
\end{equation}
This objective encourages the generator to cover diverse successful patterns, which is essential for learning transferable capability factors rather than overfitting to a narrow template.

\noindent
\textbf{Within-task Contrastive Separation.}
To explicitly suppress known failure modes for the \emph{same} task, we add a group-wise contrastive loss:
\begin{equation}
\mathcal{L}_{\mathrm{NCE}}
=
-\sum_{i}\frac{1}{|\mathcal{P}_i|}
\sum_{W^+\in\mathcal{P}_i}
\log
\frac{\exp(\ell(W^+,x(q_i))/\tau)}
{\sum_{W\in\mathcal{P}_i\cup\mathcal{N}_i}\exp(\ell(W,x(q_i))/\tau)},
\label{eq:groupnce}
\end{equation}
where $\tau$ is a temperature.
Unlike pairwise preference objectives that operate on isolated (prompt, chosen, rejected) comparisons, Eq.~\eqref{eq:groupnce} performs \emph{task-local}, \emph{listwise} discrimination.
It contrasts multiple successful workflows against task-matched failures under the same semantics and constraints, yielding a stable listwise training signal.

\begin{table*}[t]
% \vspace{-5px}
\centering
\small
\caption{Comparison of workflow Solve rate and Executability (\%) across datasets under multi-, cross-, \& unseen-domain settings.}
\vspace{-10px}
\setlength{\tabcolsep}{5pt}
\renewcommand{\arraystretch}{1.15}
\resizebox{\textwidth}{!}{
\begin{tabular}{ll l | cc cc cc cc | cc}
\toprule
% \multirow{3}{*}{\textbf{Type}} & \multirow{3}{*}{\textbf{Setting}} & \multirow{3}{*}{\textbf{Methods}} &
\multicolumn{3}{l|}{\textbf{}} &
\multicolumn{2}{c}{\textbf{Reasoning}} &
\multicolumn{2}{c}{\textbf{Coding}} &
\multicolumn{2}{c}{\textbf{Math}} &
\multicolumn{2}{c}{\textbf{Science (unseen)}} &
\multicolumn{2}{c}{\textbf{Overall}} \\
\cmidrule(lr){4-5}\cmidrule(lr){6-7}\cmidrule(lr){8-9}\cmidrule(lr){10-11}\cmidrule(lr){12-13}
%  &  &  &
\textbf{Type} & \textbf{Setting} & \textbf{Methods} &
\textbf{HotpotQA} & \textbf{DROP} &
\textbf{HumanEval} & \textbf{MBPP} &
\textbf{GSM8K} & \textbf{MATH} &
\textbf{SciBench} & \textbf{GPQA} &
\textbf{Solve} & \textbf{Exec.} \\
\midrule

\multirow{5}{*}{Manual} & \multirow{5}{*}{--} & GPT-4o-mini  & 81.54 & 81.37 & 90.07 & 70.67 & 90.14 & 51.64 & 28.40 & 34.16 & 66.00 & \textemdash \\
& & CoT & 81.75 & 83.25 & 91.60 & 70.96 & 89.95 & 53.29 & 24.83 & 35.73 & 66.42 & \textemdash \\
& & CoT-SC & 82.07 & 83.28 & 92.36 & 72.72 & 90.23 & 53.70 & 27.27 & 35.95 & 67.20 & \textemdash \\
& & Self-Refine & 81.18 & 82.50 & 90.83 & 69.50 & 88.53 & 50.20 & 27.03 & 30.25 & 65.00 & \textemdash \\
& & SPP & 81.62 & 81.62 & 92.36 & 72.72 & 91.18 & 52.88 & 28.22 & 33.72 & 66.79 & \textemdash \\
\midrule

\multirow{2}{*}{Refinement} & \multirow{2}{*}{Single-domain} & ADAS (20 iter)  & 82.84 & 82.23 & 90.83 & 70.08 & 92.03 & 52.67 & 30.36 & 36.12 & 67.14 & 79.87 \\
& & AFlow (20 iter) & 85.87 & 85.75 & 93.89 & 82.99 & 93.93 & 57.40 & 41.40 & 42.18 & 72.93 & 88.13 \\
\midrule

\multirow{6}{*}{Learning} & Single-domain & ScoreFlow (5 iter) & 86.00 & 86.14 & 95.41 & 82.69 & 94.21 & 59.25 & 34.20 & 38.69 & 72.07 & 97.36 \\
& \multirow{2}{*}{Multi-domain} & ScoreFlow (5 iter) & 85.37 & 85.16 & 93.89 & 81.23 & 93.45 & 57.81 & 34.20 & 38.69 & 71.22 & 95.89 \\
& & \textbf{CapFlow (1-pass)} & \textbf{88.12} & \textbf{87.12} & \textbf{96.18} & \textbf{83.28} & \textbf{94.97} & \textbf{59.87} & \textbf{41.60} & \textbf{42.41} & \textbf{74.19} & \textbf{98.03} \\
\cmidrule(lr){2-13}
\multicolumn{3}{l|}{\textbf{}} &
\multicolumn{2}{c}{\textbf{Code,Math->Reason}} &
\multicolumn{2}{c}{\textbf{Reason,Math->Code}} &
\multicolumn{2}{c}{\textbf{Reason,Code->Math}} &
\multicolumn{2}{c}{\textbf{All->Science}} &
\multicolumn{2}{|c}{\textbf{}} \\
\cmidrule(lr){4-5}\cmidrule(lr){6-7}\cmidrule(lr){8-9}\cmidrule(lr){10-11}
& \multirow{1}{*}{Cross-domain} & \textbf{CapFlow (1-pass)} & 86.37 & 86.25 & 93.89 & 82.11 & 92.51 & 54.32 & \textbf{41.60} & \textbf{42.41} & 72.43 & 96.87 \\
\bottomrule
\end{tabular}
}
\label{tab:main_results}
\vspace{-3px}
\end{table*}

\subsection{Counterfactual Capability Attribution for Controllable Recomposition}
\label{sec:methodology_recomposition}
The objectives above train the generator to model successful workflows, but they do not by themselves guarantee that the composer learns a \emph{structured} and \emph{transferable} decision rule.
In particular, composition can become a black box~\cite{d2022underspecification}: it may correlate with superficial features in $q$ without selecting capability factors that actually contribute to success.
To address this, we introduce a counterfactual attribution mechanism that turns which capability bases matter into a trainable credit assignment signal based on marginal effects.

\noindent
\textbf{Counterfactual Capability Attribution.}
For a sample $(q,W,s)$, let $\boldsymbol{\alpha}=\alpha(q)$ be the composer output and let $\ell^{\text{main}}$ denote the length-normalized log-likelihood under $\boldsymbol{\alpha}$.
For each basis $\Delta B_k$, we form a counterfactual composition vector by masking $\alpha_k$ and renormalizing:
\begin{equation}
\boldsymbol{\alpha}^{(-k)} \;\propto\; \boldsymbol{\alpha} \odot \big( \mathbf{1} - \mathbf{e}_k \big),
\end{equation}
and compute the counterfactual score $\ell^{(-k)}$ under $\boldsymbol{\alpha}^{(-k)}$.
We define the marginal contribution of basis $\Delta B_k$ as
\begin{equation}
\Delta_k \;=\; \ell^{\text{main}} - \ell^{(-k)}.
\label{eq:delta}
\end{equation}
A large positive $\Delta_k$ indicates that removing basis $\Delta B_k$ hurts the probability of generating this workflow, suggesting that $\Delta B_k$ encodes a capability factor relevant to the success of workflow.

\noindent
\textbf{Preference-weighted Policy Gradient for Composition.}
We use the workflow outcome $s\in[0,1]$ to assign a signed supervision signal $\mathrm{sgn}(s)=2s-1\in[-1,1]$.
We then update the composer by encouraging high composition probability for bases with positive marginal contributions on successful workflows, and discouraging such composition on failed ones.
Concretely, for a set of bases $\mathcal{K}(q)$, we optimize:
\begin{equation}
\mathcal{L}_{\mathrm{CCA}}
=
-\mathbb{E}\Big[
\sum_{\Delta B_k\in\mathcal{K}(q)}
\mathrm{sgn}(s)\cdot \Delta_k \cdot \log \alpha_k(q)
\Big]
\;+\;
\lambda_{\mathrm{dead}}\,
\mathcal{L}_{\mathrm{dead}},
\label{eq:lobo}
\end{equation}
where $\mathcal{L}_{\mathrm{dead}}$ penalizes allocating probability mass to bases whose contributions remain near zero (preventing dead or redundant bases).
Eq.~\eqref{eq:lobo} instantiates a structured, counterfactual form of credit assignment: instead of relying on indirect likelihood gradients, the composer is trained to place mass on capability factors whose \emph{removal} would counterfactually reduce success.

\noindent
\textbf{Full Objective.}
Putting together, at training time, we optimize:
\begin{equation}
\min_{\Phi,\psi}\quad
\lambda_{\mathrm{MR}}\mathcal{L}_{\mathrm{MR}}
+
\lambda_{\mathrm{NCE}}\mathcal{L}_{\mathrm{NCE}}
+
\lambda_{\mathrm{CCA}}\mathcal{L}_{\mathrm{CCA}}
+
\mathcal{L}_{\mathrm{reg}}
\label{eq:fullobj}
\end{equation}
where $\Phi$ denotes all capability basis parameters and $\psi$ the composer parameters.
The regularization term $\mathcal{L}_{\mathrm{reg}}$ includes orthogonality, entropy, and temperature deviation (detailed in Appx.~\ref{ssec:methodology_regularization}).
At inference time, CapFlow performs no iterative refinement.
In Fig.~\ref{fig:methodology_figure}, given a new task $q$, we compute $\alpha(q)$ once, activate the top-$m$ capability bases across adapted layers, and decode the workflow program $W$ in a single generation pass as Eq.~\eqref{eq:workflow_generation}.
This yields an executable workflow program with both control-flow topology and customized operator prompts, directly addressing the cost explosion, instability, and cross-domain variance of workflow refinement.

\begin{figure*}[t]
    \vspace{-5px}
    \centering
    \begin{subfigure}[t]{0.49\textwidth}
        \centering
        \includegraphics[width=\linewidth]{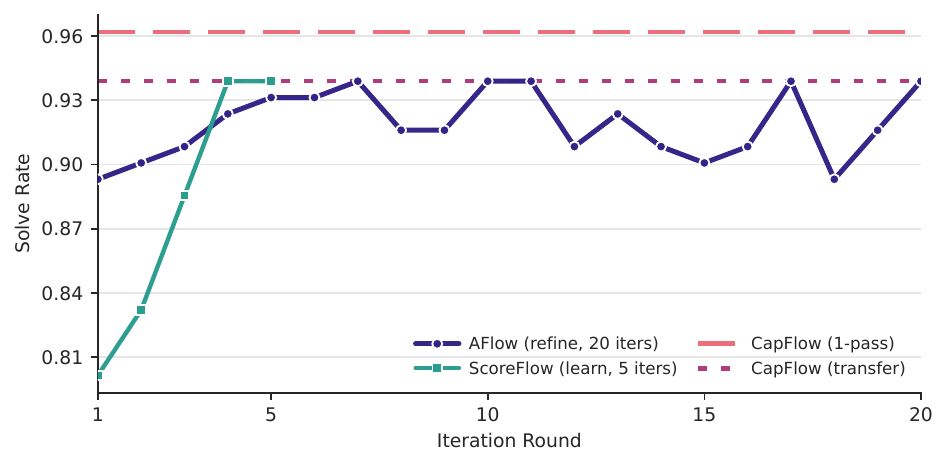}
        \vspace{-15px}
        \caption{Iteration Round vs.\ Solve Rate on HumanEval.}
        \label{fig:humaneval_iter}
    \end{subfigure}\hfill
    \begin{subfigure}[t]{0.49\textwidth}
        \centering
        \includegraphics[width=\linewidth]{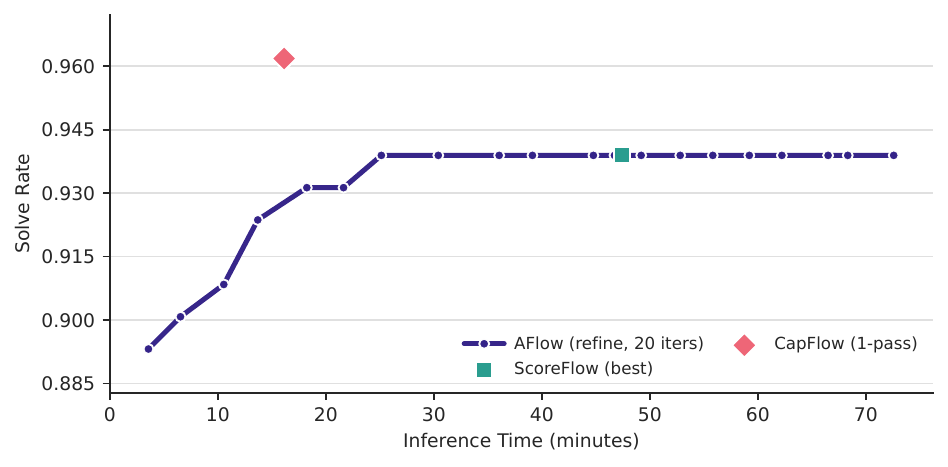}
        \vspace{-15px}
        \caption{Inference Time vs.\ Solve Rate on HumanEval.}
        \label{fig:humaneval_time}
    \end{subfigure}
    \vspace{-5px}
    \caption{Workflow generation trade-offs on HumanEval. Left: strong baselines are driven by stochastic refinement that incurs substantially higher evaluation cost for comparable gains. Right: refinement baselines improve with additional rounds but exhibit diminishing returns, whereas CapFlow achieves strong solve rates in a single generation pass.}
    \label{fig:humaneval_tradeoff}
    \vspace{-3px}
\end{figure*}

\begin{figure*}[t]
% \vspace{-5px}
\centering
\includegraphics[width=\linewidth]{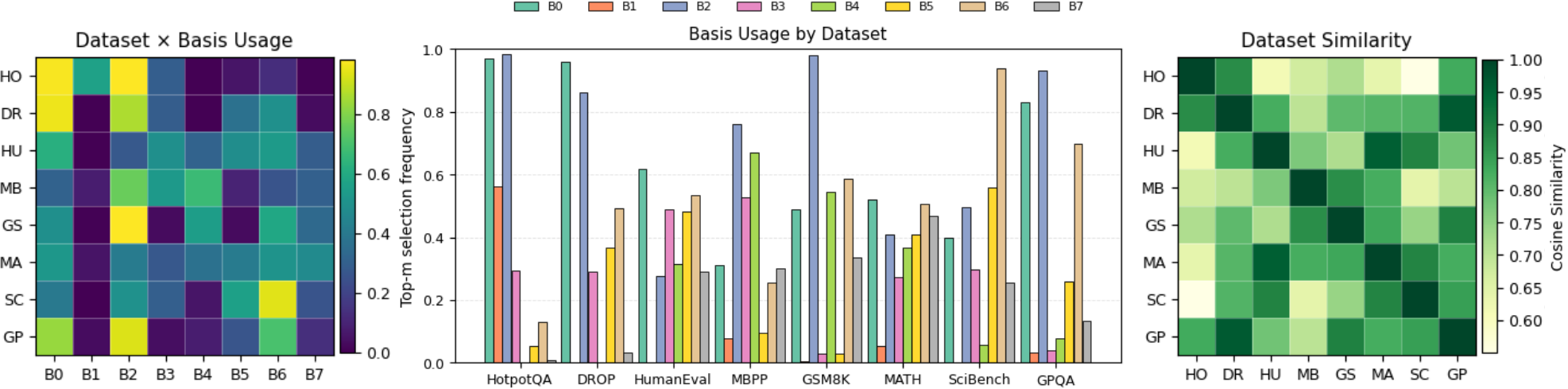}
\vspace{-15px}
\caption{Capability basis usage across domains. CapFlow maintains non-collapsed basis selection and exhibits meaningful cross-dataset overlap, supporting the intended ``reusable bases'' behavior.}
\label{fig:basis_usage_1x3}
\vspace{-3px}
\end{figure*}

\section{Experiments}
\label{sec:experiments}

\subsection{Experimental Settings}
\label{sec:exp_setup}

\noindent
\textbf{Tasks and Datasets.}
We evaluate on 8 benchmarks spanning 4 domains---reasoning~\cite{yang2018hotpotqa,dua2019drop}, coding~\cite{chen2021codex,austin2021program}, math~\cite{cobbe2021training,hendrycks2measuring}, and science~\cite{wang2024scibench,rein2024gpqa}---using established data-splitting practices~\cite{zhangaflow,yemas}.
Science is the held-out domain for \emph{unseen-domain} study. 
For the 3 training domains, the \emph{single-domain} setting involves training a separate model for each dataset and then testing it;
the \emph{multi-domain} setting trains a universal model on all reasoning, coding, and math datasets;
and the \emph{cross-domain} setting evaluates transfer via leave-one-dataset-out (LODO): for each held-out dataset, we train on the remaining datasets and test exclusively on the held-out one.

\noindent
\textbf{Baselines.}
We compare against three families of methods.
\emph{(1) Manual prompting methods} are designed by human experts, including direct invocation, Chain-of-Thought~\cite{wei2022chain}, CoT with Self-Consistency~\cite{wangself}, Self-Refine~\cite{madaan2023self}, and SPP~\cite{wang2024unleashing}.
\emph{(2) Workflow refinement methods} iteratively generate and improve workflows at inference time with a maximum budget of 20 iterations, including ADAS~\cite{huautomated} and AFlow~\cite{zhangaflow}. 
One \emph{iteration} corresponds to executing/evaluating candidate workflows on the validation set once and applying the refinement rule to produce the next candidate.
\emph{(3) Learning-based workflow generators} are trained from workflow data and generate a final workflow directly.
We consider both the \emph{single-domain} and \emph{multi-domain} model settings for ScoreFlow~\cite{wang2025scoreflow}.

\noindent
\textbf{Implementation Details.}
We employ Llama-3.2-3B-Instruct~\cite{llama32_3b_instruct,llama3herd} as the finetuning model, with Llama-3.1-8B-Instruct~\cite{llama31_8b_instruct} and Qwen2.5-Coder-3B-Instruct~\cite{qwen25coder_3b_instruct,qwen25coder_report} for comparison.
The main result is reported with 8 bases and Top-m=3, trained using 1 A100 GPU for 35 epochs at a learning rate of 2e-4 (bases) and 3e-4 (composer).
For ADAS and AFlow, we use Claude-3.5-sonnet~\cite{claude35sonnet_modelcard} as the optimizer.

\noindent
\textbf{Metrics.}
All generated workflows are executed by GPT-4o-mini with 0 temperature for a fair comparison. 
All results are averaged over 5 runs.
We report the \emph{Solve rate} (\%) for solving tasks and the \emph{Executability rate} (\%)---no runtime errors during workflow execution---on each test dataset to quantify effectiveness and stability. 
For efficiency, we additionally report the solve rate as a function of iteration and inference time to characterize the trade-off.

\subsection{Main Results}
\label{sec:exp_main}
\noindent
\textbf{Effectiveness.}
Tab.~\ref{tab:main_results} summarizes test-set results across \emph{multi-domain}, \emph{cross-domain}, and \emph{unseen-domain} settings.
Overall, CapFlow achieves the best solve rate across datasets while maintaining strong executability for reliable workflow generation.
Our 1-pass workflow generation consistently outperforms workflow refinement baselines, even when they consume 20 iteration budgets per dataset.
Notably, in the cross-domain setting, CapFlow's performance on held-out datasets can still match, and even surpass, baselines that are trained or refined on those datasets. 
This supports our claim that learning a transferable \emph{decompose--recompose--decide} mechanism yields more effective, task-specific workflow generation than heuristic trial-and-error refinement at inference time or purely imitation learning at training time.

\noindent
\textbf{Budget-Quality Trade-off.}
Workflow refinement is typically a stochastic process that is sensitive to iteration budgets.
Fig.~\ref{fig:humaneval_iter} plots the solve rate as a function of refinement iterations.
We observe diminishing returns beyond moderate budgets on AFlow: additional iterations improve performance only marginally while increasing evaluation cost at least linearly.
In contrast, CapFlow achieves competitive performance without additional evaluation cost. 

\noindent
\textbf{Inference-time Cost and Latency.}
Although refinement baselines are budgeted by iterations, iterations translate to substantial inference-time overhead due to repeated workflow executions and re-generation.
These operations are commonly priced by LLM api providers at different levels.
Beyond pricing, Fig.~\ref{fig:humaneval_time} and Tab.~\ref{tab:cost_proxy} report cost proxies, showing that CapFlow significantly reduces inference-time latency of workflow generation compared with other datasets, while achieving a higher solve rate across domains.

% \begin{figure*}[t]
% \vspace{-5px}
% \centering
% \begin{subfigure}[t]{0.346\textwidth}
%     \centering
%     \includegraphics[width=\linewidth]{figs/basis_usage_heatmap.pdf}
%     \vspace{-18px}
%     \caption{Basis usage distribution across datasets.}
%     \label{fig:basis_usage}
% \end{subfigure}
% \begin{subfigure}[t]{0.33\textwidth}
%     \centering
%     \includegraphics[width=\linewidth]{figs/dataset_similarity.pdf}
%     \vspace{-18px}
%     \caption{Dataset cosine similarity in basis usage.}
%     \label{fig:dataset_similarity}
% \end{subfigure}
% \begin{subfigure}[t]{0.297\textwidth}
%     \centering
%     \includegraphics[width=\linewidth]{figs/coactivation_heatmap.pdf}
%     \vspace{-18px}
%     \caption{Basis co-activation patterns.}
%     \label{fig:coactivation_heatmap}
% \end{subfigure}
% \vspace{-8px}
% \caption{Capability basis usage across domains. CapFlow maintains non-collapsed routing and exhibits meaningful cross-dataset overlap, supporting the intended ``reusable bases'' behavior.}
% \label{fig:basis_usage_heatmaps}
% \end{figure*}

\subsection{Capability Usage and Attribution}
\label{sec:cap_usage_attr}

We inspect whether CapFlow learns \emph{reusable} capability bases and \emph{structured} recomposition under domain shift.
Concretely, we count how often each basis is selected by the Top-$m$ composer on each dataset, and visualize the resulting bar chart and heatmaps in Fig.~\ref{fig:basis_usage_1x3}.

\noindent\textbf{Reusable Bases without Collapse.}
Fig.~\ref{fig:basis_usage_1x3} shows clear transferable usage patterns: the composer activates \emph{multiple} bases rather than collapsing to a single one, while several bases (e.g., B0, B2) are repeatedly used across different datasets.
This indicates that CapFlow captures transferable workflow factors that can be reused under domain shift, instead of learning purely dataset-specific adapters.

\noindent\textbf{Meaningful Cross-Dataset Overlap.}
To interpret overlap, we compute cosine similarity between datasets based on their basis usage frequencies (Fig.~\ref{fig:basis_usage_1x3}).
The similarity matrix is non-uniform: datasets within the same domain are generally more similar (darker diagonal), yet there remains noticeable cross-domain overlap, consistent with the intended ``reusable bases'' behavior rather than isolated per-domain routing.
This claim is consistent with our motivational study in Sec.~\ref{sec:preliminary} (Fig.~\ref{fig:workflow_analysis}).

\begin{figure}[t]
\vspace{-5px}
\centering
\includegraphics[width=0.80\linewidth]{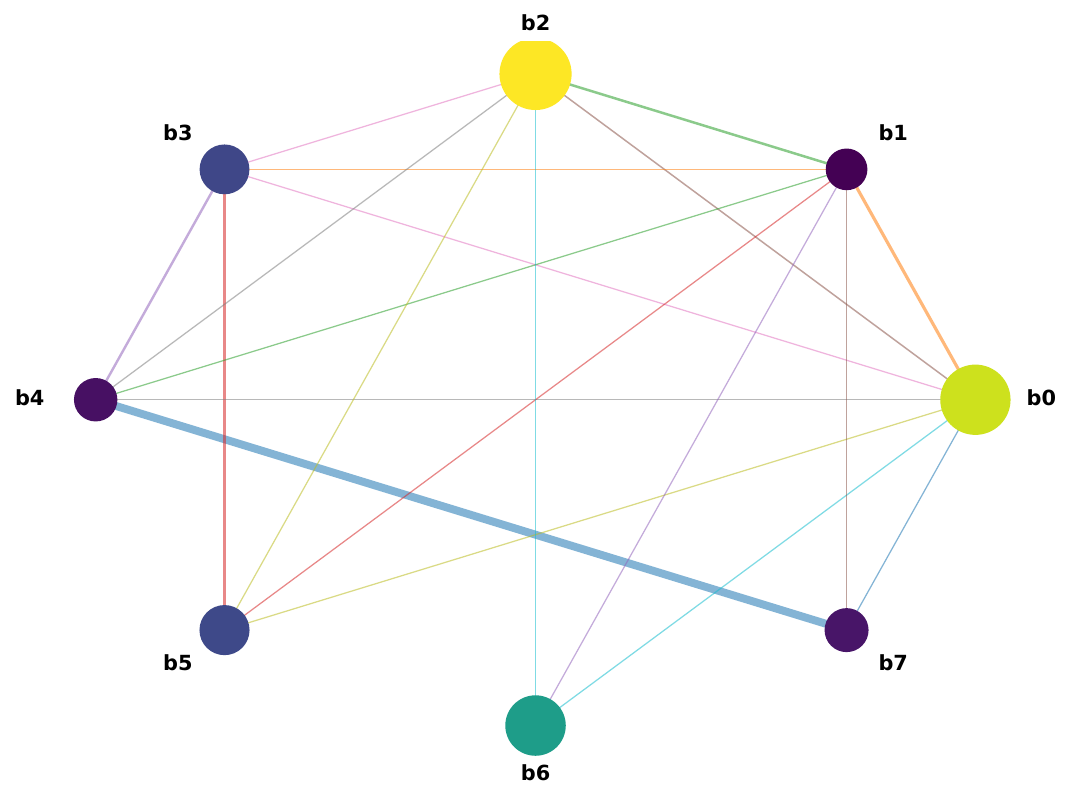}
\vspace{-10px}
\caption{Basis co-activation network. Nodes are bases. Edges show positive PMI between Top-3 basis-selection events; probabilities are estimated from empirical co-activation counts over 3706 tasks, and we visualize the top-20 pairs.}
\label{fig:pmi_network}
\vspace{-5px}
\end{figure}

\noindent\textbf{Structured Recomposition and Attribution.}
To move beyond marginal usage, we analyze \emph{synergistic} basis interactions through a positive-PMI co-activation network in Fig.~\ref{fig:pmi_network}.
PMI explicitly controls for each basis's marginal selection rate, highlighting basis pairs that are activated together \emph{more often than expected under independence}.
The resulting network is sparse and non-uniform, with a small number of consistently high-PMI edges, indicating that CapFlow's composer recombines capability bases in a \emph{coordinated} rather than independently toggled manner.
Importantly, these PMI-identified pairs provide an interpretable view of \emph{which capability combinations} the model relies on to construct effective workflows, supporting our claim that CapFlow learns \emph{structured recomposition} rather than heuristic trial-and-error routing.

\begin{table}[t]
% \vspace{-5px}
\centering
\scriptsize
\caption{Decompose ablation and cost proxies across domains. Para.: additional trainable parameter storage (MB). Train/epoch and Infer (on HumanEval) are runtimes in minutes. $K$ and Top-$m$ are different configurations of CapFlow.}
\vspace{-8px}
\setlength{\tabcolsep}{1.5pt}
\renewcommand{\arraystretch}{1.15}
\resizebox{\columnwidth}{!}{%
\begin{tabular}{lccc|cccc}
\toprule
\textbf{Variant} &
\textbf{Para. (MB)} &
\textbf{Train/epoch} &
\textbf{Infer} &
\textbf{Reas.} &
\textbf{Code} &
\textbf{Math} &
\textbf{Sci.} \\
\midrule
AFlow                 & 0.00   & 0.00     & 73.23  & 85.81 & 88.44 & 75.67 & 41.79 \\
ScoreFlow             & 13.02  & 188.40 & 47.45  & 86.07 & 89.05 & 76.73 & 36.45 \\
$K{=}1$, LoRA  & 15.28  & 19.95 & 14.17 & 87.08 & 87.95 & 76.41 & 37.47 \\
$K{=}8$, LoRAs  & 96.44 & 20.13 & 17.25 & 86.74 & 87.95 & 76.21 & 36.65 \\
$K{=}4$, Top-$m{=}3$  & 50.07  & 28.95  & 14.93  & \textbf{88.22} & 89.67 & 75.03 & 38.15 \\
$K{=}8$, Top-$m{=}3$  & 96.45  & 34.03  & 16.13  & 87.62 & \textbf{89.73} & \textbf{77.42} & \textbf{42.00} \\
$K{=}16$, Top-$m{=}6$ & 189.20 & 45.75  & 20.60  & 87.72 & 87.72 & 75.54 & 37.66 \\
$K{=}32$, Top-$m{=}9$ & 374.73 & 76.13  & 20.92  & 87.16 & 87.91 & 76.03 & 37.88 \\
\bottomrule
\end{tabular}%
}
\vspace{-5px}
\label{tab:cost_proxy}
\vspace{-3px}
\end{table}

\subsection{Ablations: Decompose-Recompose-Decide}
\label{sec:exp_ablation}

We conduct ablations aligned with our \emph{decompose-recompose-decide} mechanism to isolate which components drive performance.

\noindent
\textbf{Decompose: number of capability bases.}
Tab.~\ref{tab:cost_proxy} varies the number of capability bases $K$ to test whether a compact library of reusable capability factors is necessary.
We find that too few bases underfit diverse workflow factors and degrade transfer, while overly expanding $K$ leads to diminishing returns (and potential attribution difficulty) under domain shift.
We adopt $K=8$ and Top-$m=3$.

\noindent
\textbf{Recompose: sparse vs.\ dense composition.}
Tab.~\ref{tab:ablate_recompose} compares sparse top-$m$ composition with dense mixtures and tests sensitivity to $m$.
Sparse composition yields better controllability and transfer, while dense mixing tends to blur capability factors and increases shift variance.
Moreover, using adaptive temperature with per-task adjustment in Sec.~\ref{sec:methodology_composer} is better than standard temperature scheduling, as it captures task-specific variance across domains.

\noindent
\textbf{Decide: counterfactual contribution attribution.}
Tab.~\ref{tab:ablate_decide} evaluates the impact of the \emph{decide} component by removing counterfactual capability attribution (CCA) or group-wise contrastive loss (NCE).
CCA significantly improves the solve rate across all datasets, indicating that structured credit assignment is crucial for learning transferable capability recomposition.

\noindent\textbf{Backbone LLM.}
As a sanity check, we compare several small-scale, instruction-tuned backbones on our evaluation suite (Tab.~\ref{tab:ablate_backbone}). 
Llama-3.1-8B-Instruct and Qwen2.5-Coder-3B-Instruct are also competitive choices for LLM finetuning across domains.
We select Llama-3.2-3B-Instruct for efficiency concerns. 

% \begin{table}[t]
% \vspace{-5px}
% \centering
% \small
% \caption{\textbf{Recompose ablation.} Sparse composition improves controllability on multi- \& unseen-domains.}
% \vspace{-8px}
% \setlength{\tabcolsep}{5pt}
% \renewcommand{\arraystretch}{1.15}
% \begin{tabular}{lccc}
% \toprule
% \textbf{Variant} & \textbf{Composer} & \textbf{Solve (LODO)} & \textbf{Exec. (LODO)} \\
% \midrule
% CapFlow (full) & top-$m$ (sparse) & TBD & TBD \\
% Dense mixture & softmax (dense) & TBD & TBD \\
% Sparse (smaller $m$) & top-1 / top-2 & TBD & TBD \\
% Sparse (larger $m$)  & top-4 / top-8 & TBD & TBD \\
% \bottomrule
% \end{tabular}
% \label{tab:ablate_recompose}
% \vspace{-5px}
% \end{table}

\begin{table}[t]
\vspace{-5px}
\centering
\small
% \caption{Decompose ablation. Varying the number of capability bases. Report solve rate (\%) on multi- \& unseen-domains.}
\caption{Recompose ablation. Sparse composition improves controllability in multi-domain and unseen-domain settings.}
\vspace{-8px}
\setlength{\tabcolsep}{7pt}
\renewcommand{\arraystretch}{1.15}
\begin{tabular}{lcc|cccc}
\toprule
\textbf{Variant} & \textbf{$m$} & \textbf{Adaptive} &
\textbf{Reas.} &
\textbf{Code} &
\textbf{Math} &
\textbf{Sci.} \\
\midrule
% LoRA & 1 & 1 & TBD & TBD & TBD & TBD \\
% LoRAs & 8 & 8 & TBD & TBD & TBD & TBD \\
% CapFlow & 4 & 3 & \textbf{88.22} & 89.67 & 75.03 & 38.15 \\
Single & 1 & $\checkmark$ & 86.61 & 84.7 & 74.31 & 39.68 \\
Sparse & 2 & $\checkmark$ & 87.12 & 88.08 & 75.82 & 39.13 \\
Sparse & 3 & $\checkmark$ & \textbf{87.62} & \textbf{89.73} & 77.42 & \textbf{42.00} \\
Sparse & 3 &  & 86.47 & 88.49 & \textbf{77.78} & 36.27 \\
Dense & 8 & $\checkmark$ & 86.71 & 87.28 & 77.58 & 33.48 \\
% CapFlow & 16 & 6 & 87.72 & 87.72 & 75.54 & 37.66 \\
% CapFlow & 32 & 9 & 87.16 & 87.91 & 76.03 & 37.88 \\
\bottomrule
\end{tabular}
\label{tab:ablate_recompose}
% \vspace{-5px}
\end{table}

\begin{table}[t]
% \vspace{-5px}
\centering
\small
\caption{Decide ablation. Counterfactual attribution improves transfer and stabilizes recomposition under domain shift.}
\vspace{-10px}
\setlength{\tabcolsep}{4.9pt}
\renewcommand{\arraystretch}{1.15}
\begin{tabular}{lc|cccc}
\toprule
\textbf{Variant} &
\textbf{Decide Signal} &
\textbf{Reas.} &
\textbf{Code} &
\textbf{Math} &
\textbf{Sci.} \\
\midrule
CapFlow (Full) & CCA + preference & \textbf{87.62} & \textbf{89.73} & \textbf{77.42} & \textbf{42.00} \\
w/o NCE & CCA + imitation & 86.54 & 86.94 & 75.90 & 40.44 \\
w/o CCA & preference & 85.71 & 86.34 & 74.56 & 36.44 \\
SFT & imitation & 86.97 & 85.52 & 76.63 & 35.65 \\
\bottomrule
\end{tabular}%
\vspace{-5px}
\label{tab:ablate_decide}
% \vspace{-5px}
\end{table}

\begin{table}[t]
% \vspace{-5px}
\centering
\small
\caption{Backbone LLM ablation.}
\vspace{-10px}
\setlength{\tabcolsep}{7.5pt}
\renewcommand{\arraystretch}{1.15}
\begin{tabular}{l|cccc}
\toprule
\textbf{Backbone} &
\textbf{Reas.} &
\textbf{Code} &
\textbf{Math} &
\textbf{Sci.} \\
\midrule
Llama-3.2-3B-Instruct & 87.62 & \textbf{89.73} & \textbf{77.42} & \textbf{42.00} \\
Llama-3.1-8B-Instruct & 86.65 & 88.70 & 76.13 & 40.55 \\
Qwen2.5-Coder-3B-Instruct & \textbf{87.90} & 87.29 & 76.84 & 41.70 \\
\bottomrule
\end{tabular}%
\vspace{-5px}
\label{tab:ablate_backbone}
% \vspace{-5px}
\end{table}

\section{Conclusion}
\label{sec:conclusion}
This work addresses a central efficiency and transferability bottleneck in agentic workflow generation under domain shift.
We present \emph{Workflow Capability Basis Learning}, which internalizes a \emph{decompose-recompose-decide} mechanism into an open-source LLM to generate executable, task-specific workflows in a single pass.
Our CapFlow \emph{decomposes} workflow patterns into a compact set of reusable capability bases, \emph{recomposes} them via a task-conditioned sparse composer for new tasks, and \emph{decides} with counterfactual attribution that aligns basis selection with capabilities that truly drive workflow success, leveraging successful and failed workflows as preference supervision beyond imitation.
Across multi-domain, cross-domain, and unseen-domain evaluations, our 1-pass generator surpasses strong 20-iteration refinement baselines while substantially reducing inference-time cost and maintaining high reliability.
Future work includes extending to evolving operator libraries and tool availability, incorporating continual adaptation from execution feedback, and developing stronger diagnostics and theory for generalizing the learned capability factors to longer-horizon, more complex workflow topologies.

\clearpage
\newpage

% \begin{acks}
% \end{acks}

\bibliographystyle{ACM-Reference-Format}
\bibliography{refs}

\clearpage
\appendix

\section{Appendix}
\label{sec:appendix}

\subsection{Training and Regularization Details}
\label{sec:appendix_training_details}

This appendix provides implementation-level details for the regularization terms in $\mathcal{L}_{\mathrm{reg}}$ and the overall training procedure.

\subsubsection{Two-timescale optimization via detached routing}
A practical stability detail is to decouple how the generator (capability bases) and the composer (routing) receive gradients.
During workflow generation, we inject a \emph{detached} snapshot of routing weights into all adapted layers, i.e., $\tilde{\alpha}(q)=\mathrm{stopgrad}(\alpha(q))$.
This prevents the composer from receiving high-variance token-level gradients through every adapted module.
Instead, the composer is primarily optimized by counterfactual attribution (Eq.~\eqref{eq:lobo}) plus lightweight routing regularizers, while the capability bases are optimized by $\mathcal{L}_{\mathrm{MR}}$ and $\mathcal{L}_{\mathrm{NCE}}$.

\subsubsection{Regularization}
\label{ssec:methodology_regularization}
We add lightweight regularizers to prevent collapse and encourage a compositional capability library.
Consistent with the main paper, $\mathcal{L}_{\mathrm{reg}}$ includes (i) basis orthogonality, (ii) routing entropy, and (iii) basis dropout.

% \noindent
% \textbf{(i) Basis orthogonality (diversity across bases).}
% For each adapted layer, we encourage different bases to encode distinct directions by penalizing correlations between flattened basis parameters.
% Let $W \in \mathbb{R}^{K \times D}$ denote a basis-parameter tensor (e.g., the stacked low-rank factors) flattened into $K$ basis vectors.
% We impose:
% \begin{equation}
% \mathcal{L}_{\mathrm{ortho}}
% =
% \sum_{\text{adapted layers}}
% \sum_{W\in\{U,V\}}
% \left\|
% WW^\top - I
% \right\|_F^2,
% \label{eq:ortho}
% \end{equation}
% where the flattening is performed per basis and $I$ is the $K\times K$ identity.
% This discourages redundant bases and mitigates basis collapse.
\noindent
\textbf{(i) Basis orthogonality (diversity across bases).}
For each adapted layer $\ell$, we encourage different bases to encode distinct directions by penalizing correlations between basis-specific parameter vectors.
Let $\mathbf{B}^{(\ell)}_{U}\in\mathbb{R}^{K\times D_U}$ and $\mathbf{B}^{(\ell)}_{V}\in\mathbb{R}^{K\times D_V}$ denote the stacked-and-flattened LoRA factors across bases, where the $k$-th row is the flattened parameter vector of basis $k$ at layer $\ell$.
We impose:
\begin{equation}
\mathcal{L}_{\mathrm{ortho}}
=
\sum_{\ell\in\mathcal{L}}
\left(
\left\|\mathbf{B}^{(\ell)}_{U}\mathbf{B}^{(\ell)}_{U}{}^{\top}-I\right\|_F^2
+
\left\|\mathbf{B}^{(\ell)}_{V}\mathbf{B}^{(\ell)}_{V}{}^{\top}-I\right\|_F^2
\right),
\label{eq:ortho}
\end{equation}
where $I$ is the $K\times K$ identity.

\noindent
\textbf{(ii) Routing entropy (avoid premature collapse).}
We regularize routing entropy to encourage exploration early in training:
\begin{equation}
\mathcal{L}_{\mathrm{ent}}
=
-\lambda_{\mathrm{ent}}\,
\mathbb{E}_{q}\!\left[H(\alpha(q))\right],
\qquad
H(\alpha)=-\sum_{k=1}^{K}\alpha_k\log\alpha_k.
\label{eq:entropy_reg}
\end{equation}
Minimizing $\mathcal{L}_{\mathrm{ent}}$ increases entropy and prevents the composer from collapsing to a single basis too early.
In practice, $\lambda_{\mathrm{ent}}$ is scheduled to transition from exploration to specialization.

\noindent
\textbf{(iii) Basis dropout (compositional robustness).}
We apply \emph{basis dropout} to $\alpha(q)$ during training by randomly masking routing weights while always keeping the argmax basis active, followed by renormalization:
\begin{equation}
\begin{aligned}
&\tilde{\alpha}(q) \propto \alpha(q) \odot m(q), \\
&m_k(q) \sim \mathrm{Bernoulli}(1-p_{\mathrm{drop}}), \\
&m_{\arg\max_k \alpha_k(q)}(q) = 1,
\end{aligned}
\label{eq:basis_dropout}
\end{equation}
and $\tilde{\alpha}(q)$ is renormalized to sum to $1$.
This forces the generator to remain functional under missing capability factors and reduces single-basis dominance.

\noindent
\textbf{Optional stabilizers.}
Our implementation additionally includes two stabilizers that can be enabled at early exploration without changing the core method.
First, we encourage global utilization balance by matching the batch-mean routing distribution to uniform using a Jensen-Shannon penalty:
\begin{equation}
\mathcal{L}_{\mathrm{bal}}
=
\lambda_{\mathrm{bal}}\,
\mathrm{JS}\!\left(
\mathbb{E}_{\text{batch}}[\alpha(q)]\,\middle\|\,\mathbf{u}
\right),
\label{eq:balance_reg}
\end{equation}
where $\mathbf{u}$ is the uniform distribution over $K$ bases.
Second, when using adaptive temperature (Sec.~\ref{sec:methodology_composer}), we use it to regularize extreme temperature variation.

\subsubsection{Counterfactual attribution: implementation details}
\label{sec:appendix_counterfactual_details}
We instantiate Eq.~\eqref{eq:lobo} using counterfactual routing ablations.
For each basis $k$, we form $\alpha^{(-k)}$ by setting $\alpha_k$ to $0$ and renormalizing, then recompute the log-likelihood $\ell^{(-k)}$ and the marginal effect $\Delta_k=\ell^{\text{main}}-\ell^{(-k)}$ (Eq.~\eqref{eq:delta}).
To reduce variance, we apply per-example centering and clipping to $\{\Delta_k\}$ before using them as credit signals.

\noindent
\textbf{Dead-basis penalty.}
To operationalize $\mathcal{L}_{\mathrm{dead}}$ in Eq.~\eqref{eq:lobo}, we penalize allocating probability mass to bases whose marginal effects remain near zero:
\begin{equation}
\mathcal{L}_{\mathrm{dead}}
=
\mathbb{E}\!\left[
\sum_{k\in\mathcal{K}(q)}
\alpha_k(q)\cdot
\frac{\max(0,\gamma-|\Delta_k|)}{\gamma}
\right],
\label{eq:dead_penalty}
\end{equation}
where $\gamma$ is a margin threshold. Intuitively, if removing a basis does not change likelihood ($|\Delta_k|\approx 0$), then allocating routing mass to it is discouraged.

\noindent
\textbf{Computational knobs.}
Since counterfactual attribution requires additional forward passes, we compute it every $E$ steps and optionally on a subsampled set of examples in a batch.
In our implementation, we can attribute over all $K$ bases, while each counterfactual forward still activates only top-$m$ bases to control cost.

\subsection{Training Procedure}
\label{sec:methodology_algorithm}
Algorithm~\ref{alg:wcbl} summarizes the overall training procedure.

\begin{algorithm}[t]
\caption{Workflow Capability Basis Learning (Training)}
\label{alg:wcbl}
\begin{algorithmic}[1]
\REQUIRE Dataset $\mathcal{D}=\{(q_i,\mathcal{P}_i,\mathcal{N}_i)\}$; frozen base LLM $f_{\theta_0}$; capability bases $\Phi$; composer $g_\psi$; sparsity top-$m$; attribution interval $E$.
\ENSURE Trained $\Phi,\psi$.
\STATE Initialize $\Phi,\psi$; freeze $\theta_0$.
\FOR{each training step}
    \STATE Sample a \textbf{task-local} mini-batch from one task id, containing both successes $\mathcal{P}_i$ and failures $\mathcal{N}_i$.
    \STATE Compute task embedding $z(q)$ and routing weights $\alpha(q)=g_\psi(z(q))$.
    \STATE Apply basis dropout (Eq.~\eqref{eq:basis_dropout}); set $\tilde{\alpha}(q)=\mathrm{stopgrad}(\alpha(q))$ for adapter injection.
    \STATE Select top-$m$ bases from $\tilde{\alpha}(q)$ and broadcast to all adapted layers.
    \STATE \textbf{Update capability bases $\Phi$:} compute $\mathcal{L}_{\mathrm{MR}}$ and $\mathcal{L}_{\mathrm{NCE}}$, add $\mathcal{L}_{\mathrm{ortho}}$ (Eq.~\eqref{eq:ortho}), and take a gradient step on $\Phi$.
    \IF{step $\bmod$ $E = 0$}
        \STATE \textbf{Counterfactual attribution:} for bases $k\in\mathcal{K}(q)$, compute $\Delta_k$ via Eq.~\eqref{eq:delta} using counterfactual routings $\alpha^{(-k)}$.
        \STATE \textbf{Update composer $\psi$:} minimize $\mathcal{L}_{\mathrm{CCA}}$ (Eq.~\eqref{eq:lobo}) plus routing regularizers (Eq.~\eqref{eq:entropy_reg} and Eq.~\eqref{eq:balance_reg} temperature regularization).
    \ENDIF
\ENDFOR
\end{algorithmic}
\end{algorithm}

\end{document}